
\input harvmac

\Title{\vbox{\baselineskip12pt\hbox{CERN-TH/96-48}
\hbox{IASSNS-HEP-96/18, RI-1-96}\hbox{hep-th/9603051}}}
{Massless Monopoles Via Confining Phase Superpotentials}
\centerline{S. Elitzur $^a$ \footnote{$^1$}{e-mail address: 
elitzur@vms.huji.ac.il}, A. Forge $^a$ \footnote{$^2$} {e-mail
address: forge@vms.huji.ac.il}, A. Giveon $^b$ \footnote{$^3$}{On
leave of absence from Racah Institute of Physics, The Hebrew
University, Jerusalem 91904, Israel; e-mail address:
giveon@vxcern.cern.ch}, K. Intriligator $^c$ \footnote{$^4$}{e-mail
address keni@sns.ias.edu}, E. Rabinovici $^a$ \footnote{$^5$} {e-mail
address: eliezer@vms.huji.ac.il}}
\bigskip\centerline{$a$ Racah Institute of Physics,
The Hebrew University}
\centerline{Jerusalem, 91904, Israel}
\vskip .1in
\centerline{$b$ Theory Division, CERN, CH 1211, Geneva 23, Switzerland}
\vskip .1in
\centerline{$c$ Institute for Advanced Study, Princeton, NJ 08540, USA}

\vskip .3in

We discuss how the structure of massless monopoles in supersymmetric
theories with a Coulomb phase can be obtained from effective
superpotentials for a phase with a confined photon.  
To illustrate the technique, we derive effective superpotentials
which can be used to derive the curves which describe the Coulomb
phase of $N=2$, $SU(N_c)$ gauge theory with $N_f<N_c$ flavors. 

\Date{2/96}

\def\ev#1{\langle #1\rangle}
\def\np#1#2#3{{ Nucl. Phys.} {\bf B#1} (#2) #3}
\def\pl#1#2#3{{ Phys. Lett.} {\bf #1B} (#2) #3}

\nref\sw{N. Seiberg and E. Witten, hep-th/9407087,
\np{426}{1994}{19}; hep-th/9408099, \np{431}{1994}{484.}}
\lref\lec{For a review, see: K. Intriligator and N. Seiberg,
              hep-th/9509066,  and references therein.}
\nref\is{
 K. Intriligator and N. Seiberg, Nucl. Phys. {\bf B431} (1994) 551.}
\nref\efgr{S. Elitzur, A. Forge, A. Giveon and E. Rabinovici,
              hep-th/9504080, Phys. Lett. {\bf B353} (1995) 79.}
\nref\efgrii{ S. Elitzur, A. Forge, A. Giveon and E. Rabinovici,
                hep-th/9509130, Nucl. Phys. {\bf B459} (1996) 160.}
\lref\efgriii{S. Elitzur, A. Forge, A. Giveon and E. Rabinovici,
                hep-th/9512140.}
\lref\ils{K. Intriligator, R.G. Leigh and N. Seiberg, Phys. Rev.
             {\bf D50} (1994) 1092.}
\lref\intin{K. Intriligator, \pl{336}{1994}{409}.}
\lref\kss{D. Kutasov, A. Schwimmer and N. Seiberg, hep-th/9510222.}

\nref\klyt{ A. Klemm, W. Lerche, S. Yankielowicz and S. Theisen,
hep-th/9411048, Phys.
             Lett. {\bf B344} (1995) 169.}
\nref\af{P.C. Argyres and
             A.E. Faraggi, hep-th/9411057,
Phys. Rev. Lett. {\bf 74} (1995) 3931. }
\nref\ds{M. R. Douglas and S. H. Shenker, hep-th/9503163,
\np{447}{1995}{271}.}
\nref\ds{U.H. Danielsson and B. Sundborg, hep-th/9504102,
\pl{358}{1995}{273}.}
\nref\ad{P.C. Argyres and M. R. Douglas, hep-th/9505062,
\np{448}{1995}{93}.}
\nref\ho{ A. Hanany and Y. Oz, hep-th/9505075, 
Nucl. Phys. {\bf B452} (1995) 283.}
\nref\aps{P. Argyres, M.R. Plesser and A. Shapere, hep-th/9505100,
Phys. Rev. Lett. {\bf 75} (1995) 1699. }
\nref\bl{A. Brandhuber and K. Landsteiner, hep-th/9507008,
\pl{358}{73}{1995}.}
\nref\minnem{J. Minahan and D. Nemeschansky, hep-th/9507032;
hep-th/9601059.}
\nref\mw{E. Martinec and N. Warner, hep-th/9509161.}
\nref\asi{P. Argyres and A. Shapere, hep-th/9509175.}
\nref\hanany{A. Hanany, hep-th/9509176.}
\nref\dw{R. Donagi and E. Witten, hep-th/9510101.} 
\nref\dsii{U.H. Danielsson and B. Sundborg, hep-th/9511180.}
\nref\aam{M. Alishahiha, F. Ardalan and F. Mansouri, hep-th/9512005.}
\lref\brfs{ T. Banks and E. Rabinovici, 
Nucl. Phys. {\bf B160} (1979) 349;
             E. Fradkin and S. Shenker, Phys. Rev. {\bf D19} (1979)
3682.}

Exact results reveal that supersymmetric theories with a Coulomb phase
can have massless monopoles and/or dyons at strong coupling.  This was
first analyzed in the context of $N=2$ supersymmetric theories with
gauge group $SU(2)$, where it was found that the photon coupling is
given by the modulus of a curve which degenerates at the massless
monopole points \sw.  A similar situation occurs for $N=1$
supersymmetric theories with a Coulomb phase \refs{\is, \efgr,
\efgrii}.  Curves which give the structure of massless monopoles and
dyons have been conjectured for a number of different $N=2$ theories
\refs{\klyt -
\aam} and have been shown to pass a number of highly
non-trivial checks.  In this note we discuss certain effective
superpotentials which could be used to {\it derive} the locus of vacua
with massless monopoles dyons and the curves which give the Coulomb
gauge couplings.

Our starting point is a supersymmetric theory with a Coulomb phase,
which we drive to a partially confining phase by perturbing by a tree
level superpotential.  For example, we could consider an $N=2$
supersymmetric theory, which we break to $N=1$ by adding the tree
level superpotential $W_{N=1}=\sum _k g_kU_k$, where $U_k$ are the
Casimirs built out of the adjoint $\Phi$ of the vector multiplet.  At
the classical level, there are a number of vacua with different
expectation values of $\Phi$, breaking the gauge group to a subgroup
in which the adjoint $\Phi$ is generically massive.  The dynamics of a
vacuum with a classically unbroken subgroup $H\times U(1)^l$, with $H$
non-Abelian, is governed by the low energy $N=1$ theory with gauge
group $H$, which is confined or Higgsed, with a vacuum degeneracy
given by gaugino condensation or by a low energy superpotential.  The
$l$ photons remain massless and decoupled.  These phenomena must be
reproduced upon perturbing the low energy $N=2$ theory by $W_{N=1}$.

The spectrum of the low energy $N=2$ theory for a generic vacuum on
the Coulomb branch consists of the $R=rank(G)$ fields $U_k$ and their
associated photons.  Perturbing such a vacuum by $W_{N=1}$ does not
lead to a ground state.  The supersymmetric ground states obtained
upon perturbing by $W_{N=1}$ only occur where monopoles or dyons
become massless and can condense.  Consider, for example, the vicinity
near where a single monopole or dyon becomes massless.  The low energy
theory has a superpotential which is approximately given by
\eqn\mm{W=M(U_k)q\widetilde q+\sum _kg_kU_k}
and there is a supersymmetric ground state if there is a solution
$\ev{U_k}$ with
\eqn\vaccond{M(\ev{U_k})=0 \qquad\hbox{and}
\qquad \partial _kM(\ev{U_k})\ev{q\widetilde q}=-g_k.}  
In such a vacuum, the condensate $\ev{q\widetilde q}$ confines one of
the photons and $R-1$ are left massless.  These vacua must correspond
to the vacua in which the gauge group is classically broken to
$SU(2)\times U(1)^{R-1}$.  There are other vacua in which more than
one mutually local monopole condenses, confining more than one photon.
These vacua correspond to the classical vacua with an enhanced gauge
group which is larger than $SU(2)$.  In this letter we will focus on
the vacua with a single confined photon, corresponding to the
classical $SU(2)\times U(1)^ {R-1}$ vacua.

By finding the classical vacua in which the gauge group is broken to
$SU(2)\times U(1)^{R-1}$ and analyzing the quantum effects associated
with the low energy, $N=1$ supersymmetric $SU(2)$ theory, we can
determine where \vaccond\ must have a solution -- i.e which vacua have
a massless monopole or dyon.  This is sufficient information to derive
the elliptic curve for the gauge couplings.  This ``integrating in''
technique
\refs{\intin, \is} has been used to derive the elliptic curves for
theories with gauge group $SU(2)$ \refs{\is, \efgr, \efgrii}.  Aspects
of the $SU(N_c)$ case were discussed in \efgrii, which considered
on the vacua in which all of the photons are confined, corresponding
to the classical vacua in which $SU(N_c)$ is unbroken.

Consider, for example, $N=2$, $SU(3)$ Yang-Mills theory.  Perturbing
by $W_{N=1}=mu+gv$, where $u=\half \tr \Phi ^2$ and $v={1\over 3}\tr
\Phi ^3$, leads to classical vacua with $\Phi =0$, in which $SU(3)$ is
unbroken, and $\Phi ={\rm diag}(m/g,m/g, -2m/g)$, in which there is a
classically unbroken $SU(2)\times U(1)$.  We study the vacuum with
unbroken $SU(2)
\times U(1)$.  The low energy theory in this vacuum consists of  
a decoupled
$N=1$ photon multiplet along with $N=1$, $SU(2)$ Yang-Mills theory
with a scale $\Lambda _2$ which is related to the scale $\Lambda$ of
the high-energy $N=2$, $SU(3)$ theory by
\eqn\smi{\Lambda _2^6=(3m/g)^{-2}\cdot (3m)^2\Lambda ^6=g^2\Lambda ^6,}
where the first factor comes from matching $SU(3)$ to $SU(2)$ at the
scale $(m/g)-(-2m/g)=(3m/g)$ of the massive $SU(3)/SU(2)$ $W$ bosons
and the second factor comes from matching at the mass $W''(m/g)=(3m)$
of the massive adjoint.  The superpotential of the low energy theory
in this vacuum is
\eqn\wli{W_L={m^3\over g^2}\pm 2g\Lambda ^3,} 
where the first term is the tree level term $W_{N=1}$ evaluated for
$\Phi={\rm diag} (m/g,m/g,-2m/g)$ and the second term is the
contribution from gaugino condensation in the unbroken $N=1$ $SU(2)$,
with the sign reflecting the vacuum degeneracy.

The superpotential
\wli\ is certainly correct in the limit $m\gg \Lambda$ and $m/g\gg
\Lambda$, where the original theory is broken to our low energy theory
at a very high scale.  We will assume that \wli\ is exact for all
values of the parameters.  This assumption is referred to as the
assumption of vanishing $W_{\Delta}$ \intin.  In some cases, it is
possible to prove this assumption \refs{\intin, \is,
\efgr, \efgrii}.  In the case of \wli, however, we can not directly
rule out additive corrections of the form $W_{\Delta}=\sum
_{n=1}^\infty a_n(m^3/g^2)(g\Lambda/m)^{6n}$.  (In particular, the
condition that the dynamical superpotential of the high-energy theory
vanishes is ensured by a $U(1)_R$ symmetry, placing no further
constraint on $W_{\Delta}$.)  Assuming that such terms are absent, we
will see that we derive results which agree with those of
\refs{\klyt, \af}; conversely, it is possible 
to show that the hyper-elliptic curves of \refs{\klyt, \af} are
obtained only if $W_{\Delta}=0$.  Physically, the statement that
$W_{\Delta}=0$ reflects the absence of various operator mixings in the
flow to the infrared -- perhaps it is related to some sort of
integrability.  In any case, it is a necessary condition for obtaining
our results by ``integrating in.''

The superpotential \wli\ implies that there will be two vacua, each
with an unconfined photon, at
\eqn\uvvev{  \ev{u}=\partial _mW_L=3\left({m\over g}\right)
^2\qquad \ev{v}=\partial _g W_L=-2\left({m\over g}\right) ^3\pm
2\Lambda ^3.}  The vacua \uvvev\ must be solutions of \vaccond; in
particular, these vacua must parameterize the space of massless dyons
for $N=2$, $SU(3)$ Yang-Mills theory.  The vacua \uvvev\ parameterize
the singularities of the curve $y^2=(x^3-xu-v)^2-4\Lambda ^6$; we have
thus re-derived the result of \refs{\klyt, \af} (our normalization for
$\Lambda$ is that appropriate for the $\overline{DR}$ scheme).  In
exactly the present context, a qualitative argument was given in \ad\
that this curve must provide two solutions to the equations \vaccond.
It is possible to use the explicit expression in \af\ for the one-form
$\lambda$ to verify that \uvvev\ do indeed solve all equations in
\vaccond.  Again, we have derived the curve from \uvvev\ rather than
visa-versa.
 
This analysis can be directly extended, for example, to $N=2$
supersymmetric $SU(N_c)$ Yang-Mills theory.  Upon adding $W_{N=1}=\sum
_{r=2}^{N_c}g_ru_r$, with $u_r=\tr \Phi ^r/r$, the eigenvalues of
$\Phi$ are the roots of $W_{N=1}'(x)=g_{N_c}\prod
_{i=1}^{N_c-1}(x-a_i)$.  The vacua with classical $SU(2)\times
U(1)^{N_c-2}$ are those with, say, two eigenvalues equal to $a_1$ and
the rest given by $a_2, \dots a_{N_c-1}$.  The scale $\Lambda _2$ of
the low energy $N=1$, $SU(2)$ theory is related to the scale $\Lambda
$ of the original $N=2$, $SU(N_c)$ theory by
\eqn\smii{\Lambda _2^6=\prod _{i\neq 1}(a_1-a_i)^{-2}\cdot g_{N_c}^2
\prod _{i\neq 1}
(a_1-a_i)^2\Lambda ^{2N_c}=g_{N_c}^2\Lambda ^{2N_c},} where the first
factor again reflects the matching at the scale of the $SU(N_c)/SU(2)$
$W$ bosons and the second factor reflects the mass $W''(a_1)$ of the
adjoint in this vacuum.  Such matching relations appeared in
\kss.  The superpotential in the low energy theory is 
\eqn\wlii{W_L=W_{cl}(g)\pm 2g_{N_c}\Lambda ^{N_c},}
where $W_{cl}(g)$ is $W_{N=1}$ evaluated in the classical $SU(2)\times
U(1)^{N_c-2}$ vacua, and with the second term generated by gaugino
condensation in the low energy $SU(2)$ theory.  The superpotential
\wlii\ implies that the vacua with one photon confined and $N_c-2$
left unconfined are at
\eqn\urvevs{\ev{u_r}=u_r^{cl}(g)\pm 2\Lambda ^{N_c}\delta _{r,N_c}.}
Defining $P(x;u_r)=\det (x-\Phi )$, there is a solution $x^*$ to
$P=P'=0$ for $u_r=u_r^{cl}(g)$.  Because
$P(x;\ev{u_r})=P(x;u_r^{cl}(g))\pm 2\Lambda ^{N_c}$ for the $\ev{u_r}$
of \urvevs, for these vacua there is a solution $x^*$ to $P=\pm
2\Lambda ^{N_c}$ and $P'=0$.  The vacua \urvevs\ thus 
parameterize the singularities of the curve 
$y^2=\det (x-\Phi)^2-4\Lambda ^{2N_c}$, re-deriving the results of
\refs{\af, \klyt}.
As before, it is possible to show that this curve is obtained iff
$W_{\Delta}=0$.

These considerations can also be applied to $N=2$ theories with
hypermultiplets, where they can be written as superpotentials on the
branch in which a single photon is confined.  We first consider $N=1$
$SU(N_c)$ gauge theory with adjoint $\Phi$, a fundamental flavor $Q$
and $\tilde Q$, and tree-level superpotential
\eqn\wtreenfi{W_{tree}=
\lambda Q\Phi \tilde Q+m_QQ\tilde Q+\sum _r{g_r\over r}
\tr \Phi ^r;}
for $\lambda =1$ and $g_r=0$, this would be $N=2$, $SU(N_c)$ with a
fundamental hypermultiplet of mass $m_Q$.  For $\lambda \neq 0$, the
first term leads to a potential which lifts those moduli associated
with ``mixed'' gauge invariants, such as  $Q\Phi ^r\tilde Q$
or $\prod _{i=1}^{N_c}(\Phi ^{r_i}Q)$, which
involve both $\Phi$ and $Q$ or $\tilde Q$.  Because we want $\lambda
\neq 0$ to have an Abelian Coulomb phase, such mixed gauge invariants
are thus not moduli and there is no need to couple them to sources.  

For $g_r\neq 0$, we get a variety of classical vacua; again, we focus
on the vacua with unbroken $SU(2)\times U(1)^{N_c-2}$.  The low energy
theory in such vacua is $N=1$, $SU(2)$ with one flavor and $N_c-2$
decoupled photons.  As in
\smii, the scale $\Lambda _2$ of the low energy $SU(2)$ theory is
related to the scale $\Lambda$ of the original high energy theory by
$\Lambda _2^5=g_{N_c}^2\Lambda ^{2N_c-1}$.  The superpotential of this
low energy theory is
\eqn\wlnfi{W_L(\lambda , g,X)=W_{cl}(\lambda, g,X)+{g_{N_c}^2\Lambda
^{2N_c-1}\over X},} where $W_{cl}(\lambda , g,X)$ is the classical
superpotential \wtreenfi\ evaluated in this vacuum, with $X=Q\tilde
Q$, and the second term is the dynamically generated superpotential
for $N=1$, $SU(2)$ with one flavor.  Again, we are assuming that
$W_{\Delta}=0$.  By performing a Legendre transform {}from $\lambda$
to the conjugate variable $Z=Q\Phi \tilde Q$ and {}from the $g_r$ to
the conjugate variables $u_r$, we obtain the superpotential
\eqn\nfiwi{W=-{X\over 4\Lambda ^{2N_c-1}}P^2(x=ZX^{-1};u_r)+\lambda Z+
m_QX+\sum _rg_ru_r,}
where $P(x;u_r)=\det (x-\Phi)$.

The quantum vacua with one photon confined and $N_c-2$ left unconfined 
are given by the vacua of the superpotential \nfiwi.  These vacua have
\eqn\quanteom{P^2=4\Lambda ^{2N_c-1}(m_Q +\lambda x),\qquad 
P{\partial P\over \partial x}=2\lambda \Lambda ^{2N_c-1},} along with
some additional equations, which combine to give two vacua for each
set of the parameters.  We see from
\quanteom\ that these vacua lie on the singularity manifold of the 
hyper-elliptic curve 
\eqn\nficurv{
y^2=P^2-4\Lambda ^{2N_c-1}(m_Q+\lambda x).}  By varying the parameters
$g_r$, the vacua obtained above span the entire singularity manifold
of the curve \nficurv.  The overall scale of the $g_r$ is the order
parameter for confinement. For any value of this overall scale, the
projective space of the ratios spans the singularity manifold of the
curve. Taking the overall scale to zero while holding the ratios
fixed, we approach the transition points from the confining to the
Coulomb phase\foot{Although the original theory is $N=2$
supersymmetric only for $\lambda =1$, there is a Coulomb phase
described by the curve \nficurv\ for arbitrary $\lambda$.}.  Because
these vacua are solutions of \vaccond, the massless dyons of the
original theory occur on the singularity submanifold of the curve
\nficurv.  We thus {\it derive} the curve \nficurv\ for the photon
coupling of the original theory via the superpotential \nfiwi\ of the
branch with a single confined photon and find agreement with the
curves found in \refs{\ho, \aps}.

The generalization of \nfiwi\ to the branch with a single confined
photon for $N=1$, $SU(N_c)$ with adjoint $\Phi$ and $N_f<N_c$ flavors
is similarly found to be
\eqn\gennfw{W=-N_f\left( {\det_{N_f} X \over 4\Lambda^{2N_c-N_f}}
[\tr_{N_f} P_{N_c}(x;u_r)]^2\right)^{{1/N_f}}+
\sum_{r=2}^{N_c} g_r u_r + \tr_{N_f} mX + \tr_{N_f} \lambda Z,}
with the $N_f-1$ constraints:
\eqn\constraints{
\tr_{n}P_{N_c-N_f+n}(x;u_r-\tr_{N_f-n}x^r/r)=0, \qquad n=1,... N_f-1 .}
In these expressions, $X_i^{\tilde j}=Q_i\tilde Q^{\tilde j}$,
$Z_i^{\tilde j}=Q_i
\Phi \tilde Q^{\tilde j}$, $x\equiv ZX^{-1}$ is an $N_f\times N_f$
matrix, and $P_m(x;v_r=\tr M^r/r)\equiv \det _m(x1_{m\times m}-M)$ 
and $\tr_{N_f}=\tr_{n}+\tr_{N_f-n}$.
This result can be used to study the vacuum structure of the theory.

The technique of using an effective superpotential in the phase with a
confined photon can be applied to determine the monopole structure of
any theory with a Coulomb phase.  However, there can be subtleties
which make it more difficult than the above examples to obtain the
exact effective superpotential.  For example, as discussed in
\ref\isson{K.
Intriligator and N. Seiberg, hep-th/9503179,\np{444}{1995}{125}.},
when the index of the embedding of the unbroken $SU(2)$ in the
original gauge group is larger than one, instantons in the broken part
of the gauge group can contribute to the effective superpotential.
Also, it is possible that $W_{\Delta}\neq 0$ in some cases.
A better
understanding of such terms is needed in order to use the confining
phase superpotential technique to obtain the monopole structure of
other theories.

\bigskip
\centerline{{\bf Acknowledgments}}
AG and KI thank the ITP, Santa Barbara, where part of this work was
completed.  This research was supported in part by the NSF grant
PHY94-07194.  The work of SE is supported in part by the BRF - the
Basic Research Foundation.  The work of AG is supported in part by BSF
- American-Israel Bi-National Science Foundation, and by the BRF.  The
work of KI is supported by NSF grant PHY-9513835 and the W.M. Keck
Foundation.  The work of ER is supported in part by BSF and by the
BRF.

\listrefs
 
\end